\documentclass[aps,pre,showpacs,showkeys,twocolumn,groupedaddress]{revtex4}
\usepackage{amsmath}
\usepackage{amssymb}
\usepackage{graphicx}

\usepackage{amsfonts,amsmath}
\usepackage{graphicx}
\usepackage{mathbbol}
\newcommand {\Str} {{\rm Str}}
\newcommand {\tr} {{\rm tr}}
\newcommand {\Sdet} {{\rm Sdet}}
\newcommand {\diag} {{\rm diag}}
\newcommand {\beq}{\begin{equation}} 
\newcommand {\eeq}{\end{equation}}
\newcommand {\beqar}{\begin{eqnarray}}
\newcommand {\eeqar}{\end{eqnarray}}

\begin{document}

\title{Fidelity and level correlations in the transition from regularity to chaos}

\author{H. Kohler$^{1}$\ and C. Recher $^2$ }
\email{hkohler@icmm.csic.es}

\affiliation{ $^1$ Instituto de Ciencia de Materiales de Madrid, CSIC,
Sor Juana de la Cruz 3, Cantoblanco, 28049 Madrid, Spain\\
$^4$ Fakult\"at f\"ur Physik,
Universit\"at Duisburg--Essen,Lotharstrasse 1--21, D-47057, Germany
}

\date{\today}

\begin{abstract}Mean fidelity amplitude and parametric energy--energy correlations are calculated exactly for a regular system, which is subject to a chaotic random perturbation. It turns out that in this particular case under the average both quantities are identical. The result is compared with the susceptibility of chaotic systems against random perturbations. Regular systems are more susceptible against random perturbations than chaotic ones.
\end{abstract}


%
%
%
%
%
%

\maketitle

\section{Introduction}
A.~Peres wrote in 1984 a highly influential paper on the stability of quantum wave functions under 
random fluctuations of  the Hamiltonian \cite{per84}. He introduced the overlap between the 
wave--function propagated by  a known Hamiltonian with the same wave function propagated by a slightly 
perturbed one as a measure for stability. This quantity is nowadays known as fidelity amplitude. Its 
modulus square is called fidelity or Loschmidt echo\cite{cer70}.   Based on heuristic arguments, 
numerics and physical intuition, he concluded that a regular fluctuation, i.~e.~a fluctuations 
obeying the same superselection rules as the original Hamiltonian should have less effect on the 
stability of the quantum state than a chaotic one, i.~e. a perturbation with no additional symmetries.

Later the same question was addressed amongst others \cite{gor06,jaq08} using ensemble theory 
\cite{gor04a}.  The perturbation as well as the original Hamiltonian were chosen from a Gaussian 
random matrix ensemble 
(RME). Average fidelity amplitude $\langle f(t)\rangle $ was calculated in second order perturbation 
theory. Exponentiating the result yielded
\begin{equation}\label{gps}
\ln \langle f(t)\rangle = - 2\pi\frac{\Gamma}{D} \left(\frac{\tau}{2} + \frac{\tau^2}{\beta} + C_{\rm corr}(\tau)
\right)+{\cal O}(\Gamma^2) \ ,
\end{equation}
where $D$ is the mean level spacing  of the original Hamiltonian,  $\tau= t /t_{\rm H}$ is time 
measured in units of Heisenberg time $t_{\rm H} = 2\pi\hbar/D$ and $\Gamma$ is the Breit--Wigner 
spreading width of an unperturbed eigenstate\cite{brei36}. The first term is recognised as Fermi's 
golden rule (FGR). The second term is due to spectral fluctuations, i.~e.~fluctuations in the 
Hamiltonian which do not affect the eigenstates (called  regular fluctuations by Peres). The last 
term is a correction term which accounts for the spectral correlations of the unperturbed system.
 

The perturbative result  (\ref{gps}) was confirmed experimentally \cite{lob03a, 
schae05a,schae05b,gor06a}. Later it was completed and extended by exact calculations 
for different choices for the unperturbed system and for the perturbation. Thereby unexpected 
features  like fidelity revival at Heisenberg time 
\cite{sto04} or fidelity freeze \cite{pro04,sto06} for purely off--diagonal perturbations were revealed. 

Fidelity is, at least when averaged over a large number of eigenstates, closely related to 
the parametric energy correlator, respectively 
to its Fourier transform, the so--called cross form--factor \cite{koh08,koh11,tan96,sim95}. The latter 
measure the susceptibility of the {\em spectrum} against fluctuations of the Hamiltonian rather than the susceptibility of the 
wave--function. A relation between fidelity and these quantities is rather surprising but 
highly welcome for experimental purposes, since  spectral 
measurements are much easier to perform than measurements of fidelity, which requires in 
principle knowledge of the  entire wave--function.

The above mentioned results were derived and are valid under the assumption that already the 
unperturbed system has chaotic dynamics.  But the case originally considered by Peres, where a 
regular system is perturbed by a chaotic admixture, is in quantum information 
devices more relevant.  There the unperturbed dynamics is usually well controlled.  It has been 
object of several  theoretical and numerical \cite{pro01,pro02a,cer03,gor04a,gio10,jac10} studies 
in the recent years.  Therefore it comes as a surprise that for this case a detailed 
ensemble theoretical analysis akin to the ones performed in Refs.\cite{sto04,koh08,koh11} is 
lacking. 

The present work aims at filling this gap. We consider the same situation as Peres did: A regular 
system is perturbed by fully chaotic fluctuations. We present exact analytic results for the averaged fidelity amplitude and 
cross form--factor in the corresponding RME. It turns out that in average both  quantities are identical!  What is 
more the exact result is identical with the one obtained in exponentiated second order perturbation 
theory (\ref{gps}). 

The result is compared with the (known) results for originally chaotic systems, which are perturbed 
by random fluctuations, confirming that a regular system is more susceptible to random fluctuations 
than chaotic ones \cite{pro01,gio10}. 


\section{Definitions and Results}
\label{mod}
Fidelity amplitude is defined by ($\hbar=1$)
\begin{equation}
f(t)\ =\ \langle \psi| e^{iHt} e^{-iH_0t}|\psi\rangle \ . \label{fidelity1}
\end{equation} 
Here $H_0$ describes the regular system  and the fluctuating Hamiltonian is given by 
\begin{equation}
\label{model}
H\ =\ H_0 + \lambda D \  V \ ,
\end{equation}
where $V$ is a chaotic admixture.  We average over the spectrum of  $H_0$ in an interval which is 
large enough to contain a large number $N$ of (unperturbed) levels but small enough such that the 
mean level spacing $D$ is constant.  The strength of the perturbation is of order $D$. This means that  the dimensionless perturbation strength $\lambda$ as well as a typical matrix element of $V$ are of  order one (see Eq.~(\ref{variance})). The parameter $\lambda$ is related to the Breit--Wigner spreading width via $\Gamma=2\pi\lambda^2 D$. 
 
Following the work by Berry and Tabor \cite{ber77} in a generic regular system, like for example a rectangular billiard, the energy levels are distributed in an intervall in the same way as independent random numbers.   Assuming ergodicity the average over the energy intervall can be replaced by an ensemble average over $N$ independent random numbers or likewise over an ensemble of $N\times N$ matrices with uncorrelated eigenvalues $E_m^{(0)}, m=1,\ldots N$  (Poissonian spectrum). 

For definiteness we assume the distribution function $w(E_m^{(0)})$ of each eigenvalue of $H_0$ to be a Gaussian with zero mean and 
variance $N/2\pi$. In a region of order $N$ around the origin the eigenvalues are uniformly distributed with mean level spacing $D=w(0)$ $= N^{-1/2}$ up to corrections of order $1/N$.  This implies  a weak form of translation invariance
\begin{equation}
\label{weak}
\int dx w(x) f(x+y) \ = \  \int dx w(x) f(x) + {\cal O}(N^{-1})
\end{equation}
for any $y = {\cal O}(N^0)$, which will be used frequently.

We choose an incoherent superposition of all eigenstates in the interval as initial state. Equation  
(\ref{fidelity1}) becomes
\begin{equation}
\label{fid}
f(t) \ = \ \frac{1}{N} \tr e^{-i H t} e^{iH_0t}\ .
\end{equation}
We define the cross form--factor as  
\begin{equation}
\label{cross}
\tilde{K}(t) \ = \ \frac{1}{N} \tr e^{-i H t} \tr e^{iH_0t} \ .
\end{equation}
It is a purely spectral quantity and contains no information about the wave--function. This definition differs from to one of \cite{koh08} by a singular contribution at $t=0$ and for $N\to \infty$. 
 
 The chaotic perturbation $V$ is chosen from a Gaussian random matrix ensemble, defined by its 
second moments
\begin{equation}
\label{variance}
\langle V_{ij}V_{kl}\rangle_V \ =\  \delta_{il}\delta_{jk} +\left(\frac{2}{\beta}-1\right)\delta_{ik}\delta_
{jl}\ .
\end{equation}  
The Dyson index $\beta$ labels the three classical ensembles \cite{meh04}.  The Gaussian unitary 
ensemble (GUE, $\beta=2$, $V$ Hermitean) is chosen when the perturbation apart from being 
chaotic  breaks time reversal 
invariance (TRI). The Gaussian orthogonal ensemble (GOE, $\beta=1$, $V$ real symmetric) 
respectively the Gaussian symplectic ensemble (GSE, $\beta=4$, $V$ Hermitean selfdual) are 
chosen if the perturbation is chaotic but time reversal invariant. The GOE applies for integer spin 
and the GSE for half--integer spin.  The variance of a typical matrix element of $V$ is of order one. 
This means that the weak translation invariance (\ref{weak}) applies and the density of states of 
$H_0$ remains unaffected by the perturbation. 
 In the following angular brackets denote the above defined averages over both $V$ and $H_0$.

We now state our main results. First,  in the limit $N\to \infty$ and for $t>0$ average fidelity 
amplitude and average cross form--factor are identical
\begin{equation}
\label{mainres1}
\left< f(t)\right> \ =\ \langle \tilde{K}(t)\rangle \ .
\end{equation}
Second, for $t>0$ the exact result for the average fidelity amplitude is for all three ensembles
\begin{equation}
\langle f(t)\rangle \ =\ e^{-4\lambda^2 \pi^2\left(\frac{\tau^2}{\beta}+\frac{\tau}{2}\right)} \ , \quad \tau \ 
= \ \frac{t}{t_{\rm H}} \ .
\label{final}
\end{equation}
This result coincides with exponentiated second order perturbation theory (\ref{gps}). 
For $t=0$, $\langle \tilde{K}(t)\rangle$ differs from $\langle f(t)\rangle$ by a $\delta(t)$--contribution. 

\begin{figure}
\includegraphics[width=\linewidth]{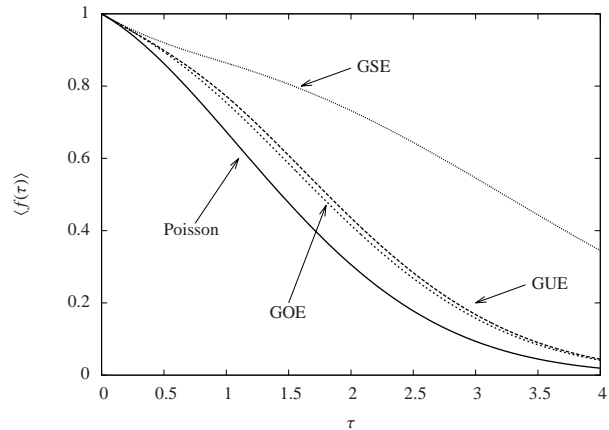}\\
  \caption{\label{fig1}Plot of equation (\ref{final}) for $\lambda=0.1$ (full line) and for a GUE 
perturbation. The curve is compared with fidelity amplitude of a chaotic original systems $H_0$, 
which is perturbed by a GUE with coupling strength 
  $\lambda=0.1$. Here $H_0$ is taken from the GUE (long dashed line, formula from \cite{sto04}), 
from the GOE (short dashed line, formula from \cite{koh11})  and from the GSE  (dotted line, formula 
from \cite{koh11}). }
\end{figure}

In Fig.~\ref{fig1} equation (\ref{final}) is plotted for a GUE pertubation ($\beta=2$). The curve is 
compared with the corresponding ones for systems which are originally chaotic and which are 
perturbed by a GUE.  We see that a regular system is less resilient against random perturbations 
than chaotic ones. This is in accordance with former results \cite{pro01,gio10} in the literature. 
Among the chaotic systems the GSE is the most resilient against random fluctuations and the GOE 
the most susceptible. This indicates a monotonous dependence of fidelity decay on the spectral 
rigidity of the original system, as quantified by the parameter $\beta$.

\section{Calculation of fidelity amplitude and cross form--factor}
\label{sec3}
Despite their simplicity, the derivation of the main identities (\ref{mainres1}) and (\ref{final}) requires 
a full fledged supersymmetric calculation. 
We sketch the main steps. 
\subsection{Map onto a supersymmetric matrix model}
After Fourier transforming (\ref{fid}) and (\ref{cross}) fidelity amplitude and cross form--factor are 
expressed in terms of the 
resolvents $G(E^\pm) = (H-E\pm i\varepsilon)^{-1}$ and $G_0(E^\pm) = (H_0-E\pm i\varepsilon)^
{-1}$ as
\begin{eqnarray}
\label{diff1}
f(t) &=& \frac{1}{N}\int \frac{d E_1 dE_2}{(2\pi)^2} e^{i (E_1-E_2) t} \tr G_0(E_1^{-})G(E_2^+)\\ 
\tilde{K}(t) &=& \frac{1}{N}\int \frac{d E_1 dE_2}{(2\pi)^2} e^{i (E_1-E_2) t} \tr G_0(E_1^{-})\tr G(E_2^
+)\ .\nonumber
\end{eqnarray}
Here and in the following we assume $t>0$ explicitely. We use the following fundamental property of the resolvent 
\begin{equation}
G_{ij}(E) \ =\ \left. \frac{1}{2}\frac{d}{d K_{ij}} \det\left(\frac{H-E+K}{H-E+K}\right)\right|_{K=0} \ ,
\end{equation}
where $K$ is a matrix containing source terms. Now both quantities can be expressed as derivatives with respect to source matrices $K^{(1)}$ and 
$K^{(2)}$ of one and the same generating function 
\begin{eqnarray}
&&Z(t, K^{(1)},K^{(2)}) \ =\   \frac{1}{N}\int \frac{d E_1 dE_2}{(2\pi)^2} e^{i (E_1-E_2) t} 
\nonumber\\
&& \ \frac{\det(H_0-E_1+K^{(1)})\det(H-E_2+K^{(2)})}{\det(H_0-E^-_1-K^{(1)})\det(H-E^+_2-K^{(2)})}
\end{eqnarray}
as
\begin{eqnarray}
\label{diff2}
f(t) &=&  \left.\sum_{n,m}\frac{\partial^2}{\partial K^{(1)}_{nm}\partial K^{(2)}_{mn}} Z\right|_{\genfrac
{}{}{0 pt}{}{K^{(1)}= 0}{K^{(2)}=0}}\cr  
\tilde{K}(t)&=& \left.\sum_{n,m} \frac{\partial^2}{\partial K^{(1)}_{nn}\partial K^{(2)}_{mm}}Z\right|_
{\genfrac{}{}{0 pt}{}{K^{(1)}= 0}{K^{(2)}=0}} \ .
\end{eqnarray}
After writing the determinants  as Gaussian integrals over vectors with commuting (denominator) 
and anticommuting (numerator) entries the average over the perturbation $\left<Z\right>_V$ can be 
performed easily.  The following standard steps are explained in detail for instance in \cite{efe97,ver85,haa01}. First a supermatrix is introduced via a Hubbard--Stratonovich 
transformation and then the differential operators in Eq.~(\ref{diff2}) are expressed as  differential operators $\Delta_f({\bf J})$ and  $\Delta_{\tilde{K}}({\bf J})$ with respect to the entries of a supermatrix ${\mathbf J}$. Finally the integrals over the vectors can be performed.

After these steps both average fidelity amplitude and average cross form--factor can be written as 
$\langle f(t)\rangle$ $=$ $\Delta_f({\bf J}) 
{\mathbf Z}({\mathbf J}, t)|_{{\mathbf J}=0}$ and 
$\langle \tilde{K}(t)\rangle$ $=$ $ \Delta_{\tilde{K}}({\bf J}) {\mathbf Z}({\mathbf J}, t)|_{{\mathbf J}=0}$, where 
the generating function $\mathbf{Z}({\mathbf J},t)$ has a supermatrix ${\mathbf J}$ as  argument. It is given by the following supersymmetric matrix integral
\begin{equation}
\label{genfunc0}
{\mathbf Z}({\mathbf J}, t)  \ =  \ \frac{1}{N}\int \frac{d E_1 dE_2}{(2\pi)^2} e^{i (E_1-E_2) t} 
\mathbf{\tilde{Z}}({\mathbf J}, 
E_1,E_2) \ ,
\end{equation}
and
\begin{eqnarray}
\label{genfunc1}
\mathbf{\tilde{Z}}({\mathbf J}, E_1,E_2)\ =\  \int d[\sigma] \exp\left(-\frac{\kappa}{2D^2\lambda^2}
\Str(\sigma - E_1)^2\right)&&\\
 \quad\left<\Sdet^{-\kappa}\left(H_0\otimes {\mathbb 1}_{4\rho} + {\mathbb 1}_N\otimes \Sigma
(\mathbf{J})\right)\right>_{H_0} \ . &&\nonumber
\end{eqnarray}
The square bracket $\langle\ldots\rangle_{H_0}$ denotes an average over the unperturbed Hamiltonian $H_0$.  We have $(\kappa,\rho) = (1/2, 2)$ for $V \in {\rm GOE}$,  $(\kappa,\rho) = (1, 1)$ for $V \in {\rm GUE}$ and $(\kappa,\rho) = (1, 2)$ for $V\in {\rm GSE}$. Here $\Sigma(\mathbf{J})$ is a $4\rho\times4\rho$ supermatrix
\begin{equation}
\Sigma({\mathbf J})\ =\ \left[\begin{array}{cc} \sigma^-& 0\\ 0&E^+_2\mathbb{1}_{2\rho}\end{array}
\right] - {\bf J}\ .
\end{equation}
 We use the standard definitions of a supertrace $\Str \sigma = \tr A_1-\tr A_2$ and of a superdeterminant $\Sdet \sigma = \det A_2^{-1}$$\det(A_1-\Lambda_1A_2^{-1}\Lambda_2)$ of a block supermatrix $\sigma = \left[\begin{array}{cc}A_1&\Lambda_1\cr\Lambda_2&A_2\end{array}\right]$, where the entries of $\Lambda_1$ and $\Lambda_2$ are anticommmuting. The $2\rho\times 2\rho$ supermatrix $\sigma$ has the form
\begin{eqnarray}
\left(\begin{array}{cc}
a_1& \lambda_1^*\cr
\lambda_1& ia_2	
\end{array}\right)&,\quad & {\rm GUE}\nonumber\\ 
\left(\begin{array}{cccc}
a_1&a_2& \lambda_1^*&-\lambda_1\cr
a_2&a_3& \lambda_2^*&-\lambda_2\cr	
\lambda_1&\lambda_2&ia_4&0\cr
\lambda_1^*&\lambda_2^*&0&ia_4
\end{array}\right)&,\quad& {\rm GOE} \nonumber\\
\left(\begin{array}{cccc}
a_1&0& \lambda_1^*&-\lambda_1\cr
0&a_1& \lambda_2^*&-\lambda_2\cr	
\lambda_1&\lambda_2&ia_2&ia_3\cr
\lambda_1^*&\lambda_2^*&ia_3&ia_4
\end{array}\right)&,\quad& {\rm GSE}\ .\label{susy}
\end{eqnarray}
The integration variables $a_n$, $n=1\ldots 4$ are real commuting variables and  $\lambda_n$, 
$n=1\ldots 2$ are complex anticommuting variables. 
The matrix integral extends over all independent entries of $\sigma$.   
The $4\rho\times 4\rho$ supermatrix 
\begin{equation}
{\mathbf J} \ =\ \left[\begin{array}{cc} J^{11}& J^{12}\\ - J^{21}& - J^{22}\end{array}\right] \ ,
\end{equation}
contains the source terms, where each of the $2\rho\times 2\rho$ matrices $J^{ij}$ has the structure 
of the prototypes (\ref{susy}).
In a tedious but straightforward calculation the operators $\Delta_f({\bf J})$ and $\Delta_{\tilde{K}}({\bf J})$ can be worked out. They  are given by
\begin{eqnarray}
\Delta_f({\bf J}) & = &\frac{1}{(2\rho)^2}\sum_n^{2\rho} \left(\sum_{m}^\rho\frac{\partial^2}{\partial J^
{21}_{nm}\partial J^{12}_{mn}}-\right.\nonumber\\
&&\qquad\qquad\left.\sum_{n=\rho+1}^{2\rho}\frac{\partial^2}{\partial J^{21}_{nm}\partial J^{12}_
{mn}}\right)\nonumber\\
\Delta_{\tilde{K}}({\bf J}) &=& \frac{1}{(2\rho)^2}\left(\sum_n^\rho\frac{\partial}{\partial J^{11}_{nn}}-
\sum_{n=\rho+1}^{2\rho}\frac{\partial}{\partial J^{11}_{nn}}\right)
\nonumber\\
&&\qquad\times\left(\sum_n^\rho\frac{\partial}{\partial J^{22}_{nn}}-\sum_{n=\rho+1}^{2\rho}\frac
{\partial}{\partial J^{22}_{nn}}\right).
\end{eqnarray}
In the following we evaluate the matrix integral (\ref{genfunc1}) in the limit $N\to
\infty$, $(E_1-E_2)/D= r$, and $r$ finite. 

\subsection{Evaluation of Eq.~(\ref{genfunc1})}
The matrix integral over  $\sigma$ is expressed in angle--eigenvalue coordinates $\sigma\to U^{-1}
sU$, $d[\sigma]\to B(s)d[s]d\mu(U)$, where $U$ denotes the supergroup, which 
diagonalizes $\sigma$.  The diagonal matrix
\begin{equation}
s \ = \ \left\{\begin{array}{lc}
 \diag(s_{{\rm B}1},is_{{\rm F}1})\ , & {\rm GUE,}\\
 \diag(s_{{\rm B}1}, s_{{\rm B}2}, is_{{\rm F}1},is_{{\rm F}2})\ , & {\rm GOE,} \\
 \diag(s_{{\rm B}1},s_{{\rm B}2} ,is_{{\rm F}1}, is_{{\rm F}2})\ , & {\rm  GSE,}\end{array}\right.
 \end{equation}
 contains the Bosonic $(s_{{\rm B}n})$ and Fermionic $(is_{{\rm F}m})$ eigenvalues.
 The Berezinian $B(s)$ is given by \cite{efe97, haa01}
\begin{equation}
B(s) \ =\ \left\{\begin{array}{cc}\displaystyle \frac{1}{(s_{{\rm B}1}-is_{{\rm F}1})^2}\ ,& {\rm GUE}\\[1em] 
\displaystyle \frac{|s_{{\rm B}1}-s_{{\rm B}2}| \delta(s_{{\rm F}1}-s_{{\rm F}2})}{(s_{{\rm B}1}- is_{{\rm 
F}1})^2 (s_{{\rm B}2}- is_{{\rm F}1})^2}\ ,& {\rm GOE}\\[1em]
\displaystyle \frac{|s_{{\rm F}1}-s_{{\rm F}2}|\delta(s_{{\rm B}1}-s_{{\rm B}2}) }{(is_{{\rm F}1}- s_{{\rm 
B}1})^2 (is_{{\rm F}2}- s_{{\rm B}1})^2}\ , & {\rm GSE .}
\end{array}\right.
\end{equation}
We proceed by observing that the average over $H_0$ in the last line of equation (\ref{genfunc1}) 
only depends on the $4\rho$ eigenvalues 
of $\Sigma({\mathbf J})$. These are only infinitesimally different from the eigenvalues of $\Sigma
(0)$. The infinitesimal corrections are obtained by a perturbative expansion. The eigenvalues of $
\Sigma(0)$ are the eigenvalues of $\sigma$ and $E_2$ which is $2\rho$--fold degenerate. We lift 
this degeneracy by replacing $E_2\mathbb{1}_{2\rho}$ with the auxiliary matrix $\mathbf{E}^
{\mathbf{a}} \ =\ \diag(E^{\mathbf{a}}_{{\rm B}1},\ldots , E^{\mathbf{a}}_{{\rm B}\rho}, E^{\mathbf{a}}_
{{\rm F}1},\ldots, E^{\mathbf{a}}_{{\rm F}\rho})$, such that $\Sigma(0)$ is completely non--degenerate.
Now the action of the operators $\Delta_f({\bf J})$ and $\Delta_{\tilde{K}}({\bf J})$ can be mapped 
in a lengthy but straightforward calculation onto the action of  first order differential operators in the eigenvalues of $s$ and ${\mathbf E}^
{\mathbf a}$. We denote them by $G^{f/\tilde{K}}$.  They read
\begin{eqnarray}
G^{f} &=& \frac{1}{(2\rho)^2}\sum_{n,m}^{\rho} \frac{{\cal D}(s_{{\rm B}n}, E^a_{{\rm B}m})+{\cal D}
(s_{{\rm B}n},- E^a_{{\rm F}m})}{s_{{\rm B}n}-E_2} \nonumber\\
                    &&        + \frac{1}{(2\rho)^2}\sum_{n,m}^{\rho} \frac{{\cal D}(is_{{\rm F}n}, E^a_{{\rm F}m})
+{\cal D}(is_{{\rm F}n}, - E^a_{{\rm B}m})}{s_{{\rm B}n}-E_2} \nonumber\\
 G^{\tilde{K}} &= &  \frac{1}{(2\rho)^2} \sum_{n,m=1}^\rho {\cal D}(s_{{\rm B}n}, -is_{{\rm F}n}){\cal D}
(E^a_{{\rm B}m}, E^a_{{\rm F}m})\ ,
 \end{eqnarray}
where we used the abbreviation ${\cal D}(x,y) =\partial/\partial x -\partial/\partial y$. 
The action of the differential operator with respect to the source matrix ${\mathbf J}$  on $\mathbf{\tilde
{Z}}({\mathbf J}, E_1,E_2)$ can be replaced by  the action of $G^{f/\tilde{K}}$ as follows
\begin{eqnarray}
\label{genfunc2}\label{eq7}
\left.\Delta_{f/\tilde{K}}({\bf J}) \mathbf{\tilde{Z}}({\mathbf J}, E_1,E_2)\right|_{{\mathbf J}=0} \ =\  \qquad\qquad
\qquad\qquad\qquad&&\\ 
\ \left. \int B(s)d[s] e^{-\frac{\kappa}{2D\lambda^2}\Str(s - E_1)^2} G^{f/\tilde{K}}Z_0(s,{\mathbf E}^
{\mathbf a})\right|_{\mathbf{E}^{\mathbf{a}}= E_2\mathbb{1}_{2\rho}} . && \nonumber
\end{eqnarray}
Here $Z_0(s,{\mathbf E}^{\mathbf a})$ is the superdeterminant in the second line of equation (\ref
{genfunc1}).  The matrices $s$  and ${\mathbf E}^{\mathbf a}$ are diagonal, thus 
the superdeterminant can be 
written as a ratio of ordinary determinants
\begin{eqnarray}
Z_0(s,{\mathbf E}^{\mathbf a})= \qquad\qquad\qquad\qquad\qquad\qquad\qquad\qquad &&
\nonumber\\
\left<\left(\frac{\prod_{n=1}^\rho\det(H_0-is_{{\rm F}n})\det(H_0-E^{\mathbf{a}}_{{\rm F}n})}{\prod_
{n=1}^\rho\det(H_0-s^-_{{\rm B}n})\det(H_0-E^{\mathbf{a}+}_{{\rm B}n})}\right)^\kappa\right> \ .&&
\end{eqnarray}
Since the eigenvalues of $H_0$ are uncorrelated the ensemble average over $H_0$ of $Z_0$ is 
the $N$--th power of a single integral 
\begin{equation}
Z_0(s,{\mathbf E}^{\mathbf a})\ =\ \left(\int dx w(x) R(x,s,{\mathbf E}^{\mathbf a}) \right)^N\ ,
\end{equation}
where $R$ is given by
\begin{equation}
R(x,s,{\mathbf E}^{\mathbf a}) \ =\ \left(\frac{\prod_{n=1}^\rho (x-is_{{\rm F}n})(x-E^{\mathbf{a}}_{{\rm 
F}n})}{\prod_{n=1}^\rho(x-s^-_{{\rm B}n})(x-E^{\mathbf{a}+}_{{\rm B}n})}\right)^\kappa \ .
\end{equation}
Next we have to calculate the action of $G^{f/\tilde{K}}$ on $Z_0$. We observe that at $\mathbf{E}^
{\mathbf{a}}= E_2\mathbb{1}_{2\rho}$ the $E_2$ dependence of $Z_0$ drops out and $Z_0$ can 
be evaluated in the large $N$--limit 
\begin{equation}
\left.\lim\limits_{N\to \infty}Z_0(s,{\mathbf E}^{\mathbf a})\right|_{\mathbf{E}^{\mathbf{a}}= 
E_2\mathbb{1}_{2\rho}} \ =\ \exp\left(-i\pi \kappa D N \Str s\right) \ .
\end{equation} 
We need  to calculate the action of $G^{f/\tilde{K}}$ on $R$. One finds in a 
straightforward calculation
\begin{eqnarray}
&&\left.  G^{f/\tilde{K}} \int dx w(x) R(x,s,{\mathbf E}^{\mathbf a}) \right|_{\mathbf{E}^{\mathbf{a}}= 
E_2\mathbb{1}_{2\rho}} \ = \nonumber\\
&&\qquad \sum_{n=1}^\rho {\cal D}(s_{{\rm B}n}, -is_{{\rm F}n}) \int dx \ w(x)\frac{\tilde{R}(s,x)}{x-
E_2^+} \ , \label{equality1}
\end{eqnarray}
where
\begin{equation}
\tilde{R}(s,x)\ =\ \left\{\begin{array}{cc} \displaystyle \sqrt{\frac{(x-is_{{\rm F}1})(x-is_{{\rm F}2})}{(x- s_{{\rm B}1}^-)(x-s_
{{\rm B}2}^-)}}& {\rm GOE}\\[1em]
     \displaystyle \frac{x- is_{{\rm F}1}}{x- s_{B1}^-} & {\rm GUE}\\[1em] 
   \displaystyle \frac{(x-is_{{\rm F}1})(x-is_{{\rm F}2})}{(x-s_{B1}^-)(x-s_{B2}^-)} & {\rm GSE.}\end{array}\right. 
\end{equation}
Equation (\ref{equality1}) holds for both $G^{f}$ and for $G^{\tilde{K}}$. From this follows 
immediately $\langle f(t)\rangle =\langle \tilde{K}(t)\rangle$,  which is our first main result (\ref{mainres1}). 

Collecting the former results and expressing everything in terms of the dimensionless energy 
difference $r$, we find 
\begin{eqnarray}\label{genfunc2aa}
\langle f(t)\rangle &=& \int \frac{d r}{(2\pi)^2} e^{2\pi i r \tau} \int d[s]  B(s) e^{-\frac{\kappa}{2\lambda^2}\Str s^2 -i\pi\kappa \Str s}\nonumber\\
&&\ \sum_{n=1}^\rho {\cal D}(s_{{\rm B}n}, -is_{{\rm F}n}) \int dx \ w(x) \frac{\tilde{R}(s,x)}{x+r^-} \ .
\end{eqnarray}
We recall $\tau$ $=t/t_{\rm H}$. Absolute 
convergence of the $x$--integration and of the $s$ integration is guaranteed by the Gaussian 
weight functions. We are therefore allowed to interchange order of integration and perform the $r$ 
integration by the residue theorem. Moreover, we perform an integration by parts of the operator $
\sum_{n=1}^\rho {\cal D}(s_{{\rm B}n}, -is_{{\rm F}n})$ using that this operator annihilates $\Str s$ 
and $B(s)$.
The result is
\begin{eqnarray}\label{genfunc2a}
\langle f(t)\rangle &=& \frac{i \kappa}{\pi \lambda^2} \int d[s] B(s) e^{-\frac{\kappa}{2\lambda^2}\Str s^2 -i\pi\kappa \Str s}\nonumber\\
&&\qquad \times\ \Str s \int dx \, w(x) e^{-2\pi i x \tau}\tilde{R}(s,x) \ .
\end{eqnarray}

The remaining integral is simple for the GUE but somewhat tricky for the GOE and for the GSE. 

In the GUE--case the $x$--integration can be performed by the residue theorem and employing 
weak translation invariance. As a result the remaining integrals over $s_{{\rm B}1}$ and over $s_
{{\rm F}1}$ decouple. Both are Gaussian integrals. The final result is
\begin{equation}
\langle f(t)\rangle_{\rm GUE} \ =\ e^{-2\lambda^2 \pi^2\left(\tau^2+\tau \right)} \ .
\end{equation}

For the GOE the $x$--integration is more complicated due to the square roots in $\tilde{R}$ and the
more complicated Berezinian $B(s)$. The expression on the r.h.s. of (\ref{genfunc2a}) is 
a fourfold integral over the three eigenvalues of $\sigma$ and over $x$. In order to simplify this 
integral, we use  identity 
(27) of Ref.~\cite{groe04} and thereby extract a GUE contribution from Eq.~(\ref{genfunc2a}). 
Thus the average fidelity amplitude is given by the GUE result plus an extra term
\begin{equation}
\langle f(t)\rangle_{\rm GOE} \ =\ \langle f(t)\rangle_{\rm GUE} +\langle f(t)\rangle_{\rm add}\ .
\end{equation}
After using identity (28) of Ref.~\cite{groe04} and an integration by parts the extra term reads
\begin{eqnarray}\label{genfunc3}
&&\langle f(t)\rangle_{\rm add} \ = \  - \frac{i}{2\pi \lambda^2} \int \frac{ds_{{\rm B}1}ds_{{\rm B}2}ds_
{{\rm F}1}ds_
{{\rm F}2}|s_{{\rm B}1}-s_{{\rm B}2}|}{(s_{{\rm B}1}- is_{{\rm F}1}) (s_{{\rm B}2}- is_{{\rm F}1})}
\nonumber\\
&& \qquad \times \delta(s_{{\rm F}1}-s_{{\rm F}2}) \exp\left(-\frac{1}{4\lambda^2}\Str s^2 -\frac{i\pi}{2}\Str s\right)\nonumber\\
&&\times\left( \sum_{n=1}^\rho {\cal D}(s_{{\rm B}n}, -is_{{\rm F}n}) - \frac{\Str s}{s_{{\rm B}1}-s_{{\rm B}2}}
{\cal D}(s_{{\rm B}1}, s_{{\rm B}2}) \right)\nonumber\\
&&\qquad\qquad \int dx \, w(x) e^{-2\pi i x \tau}\tilde{R}(s,x) \ .
\end{eqnarray}
The action of the differential operator in the third line of equation (\ref{genfunc3}) on $\tilde{R}(s,x)$ 
yields the crucial simplification of the integral 
\begin{eqnarray}\label{genfunc3a}
&&\langle f(t)\rangle_{\rm add} \ = \ \frac{i}{2\pi\lambda^2}  \int ds_{{\rm F}1}ds_{{\rm B}1}ds_{{\rm B}
2} |s_{{\rm B}1}-s_{{\rm B}2}|\nonumber\\
&&\qquad \int dx \, w(x) \frac{e^{-\frac{1}{4\lambda^2}\Str s^2 -\frac{i\pi}{2}\Str s-2\pi i x \tau}}{\sqrt{x-s_{{\rm B}1}^-}^3 \sqrt{x-s_{{\rm B}2}^-}^3}\ .
\end{eqnarray}
Now the (trivial) $s_{{\rm F}1}$ integration decouples 
from the rest.  Introducing coordinates $v= s_
{{\rm B}1}-s_{{\rm B}2}$ and $u = (s_{{\rm B}1}+s_{{\rm B}2})/2$ and using the weak translation 
invariance of $w(x)$ it is seen that the $x$--integration does not depend on $u$. Thus the $u$--integration 
decouples as well. Moreover the integrals over $s_{{\rm F}1}$ and over $u$ together 
yield again the GUE result. Thus we can write
\begin{eqnarray}\label{genfunc4}
\langle f(t)\rangle_{\rm add} &= & \frac{i}{4\pi D}\langle f(t)\rangle_{\rm GUE}  \int dv |v|\nonumber\\
&&\int dx \, w(x) \frac{e^{-\frac{v^2}{2\lambda^2}} e^{-2\pi i x \tau}}{\sqrt{x-v^-}^3 \sqrt{x+v^+}^3}\ .
\end{eqnarray} 
The $x$--integration can performed employing weak translation invariance of $w(x)$ 
\begin{equation}
\int dx \, w(x) \frac{e^{-2\pi i x \tau}}{\sqrt{x-v^-}^3 \sqrt{x+v^+}^3}\ = \ -2\pi i D \frac{d}{d v} {\rm J}_0
(2\pi \tau v)   \  .
\end{equation}
Here  ${\rm J}_0(x)$ is the zeroth order Bessel function. After yet another integration by parts
\begin{eqnarray}
&&\langle f(t)\rangle_{\rm add} =  \langle f(t)\rangle_{\rm GUE}\nonumber\\
       &&\qquad \times\left(-{\rm J}_0(0)+ \int dv |v| e^{-\frac{v^2}{2\lambda^2}} {\rm J}_0(2\pi \tau v) 
\right)\ .
\end{eqnarray}
The remaining integral over $v$ is a standard Gradsteyn integral \cite{gra00}. With ${\rm J}_0(0)
=1$ we find
\begin{equation}
\langle f(t)\rangle_{\rm add} \ = \  -\langle f(t)\rangle_{\rm GUE} \left(1- e^
{-2\pi^2\tau^2\lambda^2}\right) \  
\end{equation}
and finally
$
\langle f(t)\rangle_{\rm GOE} \ =\ e^{-4\lambda^2 \pi^2\left(\tau^2+\frac{\tau}{2}\right)} \ .
$
Again the result coincides with exponentiated second order perturbation theory. Here it is even 
more surprising than for the GUE since the deceptively simple and compact result  is 
-- at least in the way it was calculated here -- the outcome of a complicated conspiration of terms.    

The GSE case is treated like the GOE case.  The calculation is simpler since no square root 
appears in $\tilde{R}(s,x)$. The result is
$\langle f(t)\rangle_{\rm GSE} \ =\ e^{-4\lambda^2 \pi^2\left(\frac{\tau^2}{4}+\frac{\tau}{2}\right)}$.
Thus we can write the mean fidelity amplitude in all three cases concisely as in equation (\ref
{final}).

\section{Conclusion}
\label{conc}

Using supersymmetry we calculated  fidelity amplitude and cross form--factor in a random matrix model for a 
regular system with a chaotic perturbation. Surprisingly both quantities are identical under ensemble average. 
The result is a simple exponential of two terms. One term decays quadratically in time and one term decays linearly. 
Exponentiated second order perturbation theory is exact, indicating that a more direct proof of our results is likely to exist. 

The fact that both quantities are identical can be reconciled with the general 
differential identity \cite{sim95,koh11} 
\begin{eqnarray}\label{gen}
   \langle f(t)\rangle &=&-\frac{\beta}{4\pi^2\tau^2}
                \frac{\partial}{\partial (\lambda_\parallel^2)} \langle\tilde{K}(t)\rangle\ ,
\end{eqnarray}
 where only the subspace of the perturbation which is parallel to the original Hamiltonian enters.  
To understand this  we split the matrix space of perturbations ${\cal V}\ni V$ into a subspace ${\cal 
V}_\parallel$ parallel and a subspace ${\cal V}_\perp$ perpendicular to the matrix space of the unperturbed 
Hamiltonian ${\cal H}_0\ni H_0 $ via $V \in {\cal V}_\perp \Leftrightarrow$  $ \tr VH_0=0$, $\forall H_0\in{\cal H}_0$ and ${\cal V}_\parallel = {\cal V} - 
{\cal V}_\parallel$.  Likewise any perturbation can be written as $V= V_\parallel + V_\perp$, where 
the parallel part shares the symmetries of $H_0$. The model (\ref{model}) might be generalized 
to $ H = H_0 +D\lambda_{\parallel}V_\parallel + D \lambda_\perp V_\perp $ where coupling strengths $\lambda_\parallel
$ and $\lambda_\perp$ might be different.  
 
Since a chaotic perturbation breaks any symmetry of the regular system in our case $V_\parallel$ 
consists in the truncation of $V$ to its diagonal part in the eigenbasis of $H_0$.  It is intuitively clear 
and has been shown perturbatively \cite{cer03}  that the diagonal part of the perturbation in the 
eigenbasis of $H_0$ is responsible for the Gaussian decay. On the other hand the linear term 
(FGR) is due to the off-diagonal terms. This suggests to write our main results (\ref{mainres1}) and 
(\ref{final}) in the form
\begin{equation}
\langle \tilde{K}(t)\rangle \ =\ \langle f(t)\rangle \ =\ e^{-4\pi^2\left( \frac{\lambda_\parallel^2\tau^2}
{\beta}+ \frac{\tau\lambda_\perp^2}{2}\right)} \ ,
\label{final1}
\end{equation}
which obeys the differential relation (\ref{gen}). Although we have proven equation (\ref{final1}) 
only for $\lambda_\parallel=\lambda_\perp$, we conjecture it to hold exactly for arbitrary coupling 
strength $\lambda_\parallel$ and $\lambda_\perp$.  This conjecture is backed by a perturbative calculation akin to that
leading to Eq.~(\ref{gps}). A rigorous 
proof remains a challenge 
for the future.  

\acknowledgements
 HK acknowledges support from Deutsche Forschungsgemeinschaft by the
grants KO3538/1-2 and from CSIC (Spain) through the JAE program. CR acknowledges support by the " Studienstiftung des deutschen Volkes". 

\end{document}